\RequirePackage{lineno}
\documentclass[aps,amssymb,amsmath,aps,prd,floatfix,showpacs,letterpaper,twocolumn,longbibliography,superscriptaddress]{revtex4-1}
\usepackage{graphicx}
\usepackage[caption=false]{subfig}
\usepackage{bm}
\usepackage{epsf}
\usepackage{rotating}
\usepackage{epsfig,graphics,rotate,color}
\usepackage{wrapfig}
\usepackage{array,hhline,dcolumn}%
\usepackage{placeins}
\usepackage{booktabs}
\usepackage[colorlinks=true,citecolor=blue,linkcolor=blue]{hyperref}
\usepackage[utf8]{inputenc}
\usepackage{microtype}
\usepackage{siunitx}
\usepackage{stackengine}
\usepackage[T5,T1]{fontenc}

\newcommand{\nusquids}{\texttt{nuSQuIDS}}
\newcommand{\emcee}{\texttt{emcee}}
\newcommand\barparen[1]{\overset{\scriptscriptstyle(-)}{#1}}
\newcommand{\fratio}{\Phi/\Phi_0}
\newcommand{\leff}{\mathcal{L}_{\rm Eff}}

\AtBeginDocument{
\heavyrulewidth=.08em
\lightrulewidth=.05em
\cmidrulewidth=.03em
\belowrulesep=.65ex
\belowbottomsep=0pt
\aboverulesep=.4ex
\abovetopsep=0pt
\cmidrulesep=\doublerulesep
\cmidrulekern=.5em
\defaultaddspace=.5em
}

\begin{document}

\collaboration{IceCube Collaboration}
\noaffiliation

\title{Measurement of the high-energy all-flavor neutrino-nucleon cross section with IceCube}

\begin{abstract}
The flux of high-energy neutrinos passing through the Earth is attenuated due to their interactions with matter. The interaction rate is modulated by the neutrino interaction cross section and affects the flux arriving at the IceCube Neutrino Observatory, a cubic-kilometer neutrino detector embedded in the Antarctic ice sheet. We present a measurement of the neutrino cross section between \SI{60}{\tera \eV} and \SI{10}{\peta \eV} using the high-energy starting events (HESE) sample from IceCube with 7.5 years of data. The result is binned in neutrino energy and obtained using both Bayesian and frequentist statistics. We find it compatible with predictions from the Standard Model. Flavor information is explicitly included through updated morphology classifiers, proxies for the the three neutrino flavors. This is the first such measurement to use the three morphologies as observables and the first to account for neutrinos from tau decay.
\end{abstract}

\affiliation{III. Physikalisches Institut, RWTH Aachen University, D-52056 Aachen, Germany}
\affiliation{Department of Physics, University of Adelaide, Adelaide, 5005, Australia}
\affiliation{Dept. of Physics and Astronomy, University of Alaska Anchorage, 3211 Providence Dr., Anchorage, AK 99508, USA}
\affiliation{Dept. of Physics, University of Texas at Arlington, 502 Yates St., Science Hall Rm 108, Box 19059, Arlington, TX 76019, USA}
\affiliation{CTSPS, Clark-Atlanta University, Atlanta, GA 30314, USA}
\affiliation{School of Physics and Center for Relativistic Astrophysics, Georgia Institute of Technology, Atlanta, GA 30332, USA}
\affiliation{Dept. of Physics, Southern University, Baton Rouge, LA 70813, USA}
\affiliation{Dept. of Physics, University of California, Berkeley, CA 94720, USA}
\affiliation{Lawrence Berkeley National Laboratory, Berkeley, CA 94720, USA}
\affiliation{Institut f{\"u}r Physik, Humboldt-Universit{\"a}t zu Berlin, D-12489 Berlin, Germany}
\affiliation{Fakult{\"a}t f{\"u}r Physik {\&} Astronomie, Ruhr-Universit{\"a}t Bochum, D-44780 Bochum, Germany}
\affiliation{Universit{\'e} Libre de Bruxelles, Science Faculty CP230, B-1050 Brussels, Belgium}
\affiliation{Vrije Universiteit Brussel (VUB), Dienst ELEM, B-1050 Brussels, Belgium}
\affiliation{Department of Physics and Laboratory for Particle Physics and Cosmology, Harvard University, Cambridge, MA 02138, USA}
\affiliation{Dept. of Physics, Massachusetts Institute of Technology, Cambridge, MA 02139, USA}
\affiliation{Dept. of Physics and Institute for Global Prominent Research, Chiba University, Chiba 263-8522, Japan}
\affiliation{Department of Physics, Loyola University Chicago, Chicago, IL 60660, USA}
\affiliation{Dept. of Physics and Astronomy, University of Canterbury, Private Bag 4800, Christchurch, New Zealand}
\affiliation{Dept. of Physics, University of Maryland, College Park, MD 20742, USA}
\affiliation{Dept. of Astronomy, Ohio State University, Columbus, OH 43210, USA}
\affiliation{Dept. of Physics and Center for Cosmology and Astro-Particle Physics, Ohio State University, Columbus, OH 43210, USA}
\affiliation{Niels Bohr Institute, University of Copenhagen, DK-2100 Copenhagen, Denmark}
\affiliation{Dept. of Physics, TU Dortmund University, D-44221 Dortmund, Germany}
\affiliation{Dept. of Physics and Astronomy, Michigan State University, East Lansing, MI 48824, USA}
\affiliation{Dept. of Physics, University of Alberta, Edmonton, Alberta, Canada T6G 2E1}
\affiliation{Erlangen Centre for Astroparticle Physics, Friedrich-Alexander-Universit{\"a}t Erlangen-N{\"u}rnberg, D-91058 Erlangen, Germany}
\affiliation{Physik-department, Technische Universit{\"a}t M{\"u}nchen, D-85748 Garching, Germany}
\affiliation{D{\'e}partement de physique nucl{\'e}aire et corpusculaire, Universit{\'e} de Gen{\`e}ve, CH-1211 Gen{\`e}ve, Switzerland}
\affiliation{Dept. of Physics and Astronomy, University of Gent, B-9000 Gent, Belgium}
\affiliation{Dept. of Physics and Astronomy, University of California, Irvine, CA 92697, USA}
\affiliation{Karlsruhe Institute of Technology, Institute for Astroparticle Physics, D-76021 Karlsruhe, Germany }
\affiliation{Dept. of Physics and Astronomy, University of Kansas, Lawrence, KS 66045, USA}
\affiliation{SNOLAB, 1039 Regional Road 24, Creighton Mine 9, Lively, ON, Canada P3Y 1N2}
\affiliation{Department of Physics and Astronomy, UCLA, Los Angeles, CA 90095, USA}
\affiliation{Department of Physics, Mercer University, Macon, GA 31207-0001, USA}
\affiliation{Dept. of Astronomy, University of Wisconsin{\textendash}Madison, Madison, WI 53706, USA}
\affiliation{Dept. of Physics and Wisconsin IceCube Particle Astrophysics Center, University of Wisconsin{\textendash}Madison, Madison, WI 53706, USA}
\affiliation{Institute of Physics, University of Mainz, Staudinger Weg 7, D-55099 Mainz, Germany}
\affiliation{Department of Physics, Marquette University, Milwaukee, WI, 53201, USA}
\affiliation{Institut f{\"u}r Kernphysik, Westf{\"a}lische Wilhelms-Universit{\"a}t M{\"u}nster, D-48149 M{\"u}nster, Germany}
\affiliation{Bartol Research Institute and Dept. of Physics and Astronomy, University of Delaware, Newark, DE 19716, USA}
\affiliation{Dept. of Physics, Yale University, New Haven, CT 06520, USA}
\affiliation{Dept. of Physics, University of Oxford, Parks Road, Oxford OX1 3PU, UK}
\affiliation{Dept. of Physics, Drexel University, 3141 Chestnut Street, Philadelphia, PA 19104, USA}
\affiliation{Physics Department, South Dakota School of Mines and Technology, Rapid City, SD 57701, USA}
\affiliation{Dept. of Physics, University of Wisconsin, River Falls, WI 54022, USA}
\affiliation{Dept. of Physics and Astronomy, University of Rochester, Rochester, NY 14627, USA}
\affiliation{Oskar Klein Centre and Dept. of Physics, Stockholm University, SE-10691 Stockholm, Sweden}
\affiliation{Dept. of Physics and Astronomy, Stony Brook University, Stony Brook, NY 11794-3800, USA}
\affiliation{Dept. of Physics, Sungkyunkwan University, Suwon 16419, Korea}
\affiliation{Institute of Basic Science, Sungkyunkwan University, Suwon 16419, Korea}
\affiliation{Dept. of Physics and Astronomy, University of Alabama, Tuscaloosa, AL 35487, USA}
\affiliation{Dept. of Astronomy and Astrophysics, Pennsylvania State University, University Park, PA 16802, USA}
\affiliation{Dept. of Physics, Pennsylvania State University, University Park, PA 16802, USA}
\affiliation{Dept. of Physics and Astronomy, Uppsala University, Box 516, S-75120 Uppsala, Sweden}
\affiliation{Dept. of Physics, University of Wuppertal, D-42119 Wuppertal, Germany}
\affiliation{DESY, D-15738 Zeuthen, Germany}

\author{R. Abbasi}
\affiliation{Department of Physics, Loyola University Chicago, Chicago, IL 60660, USA}
\author{M. Ackermann}
\affiliation{DESY, D-15738 Zeuthen, Germany}
\author{J. Adams}
\affiliation{Dept. of Physics and Astronomy, University of Canterbury, Private Bag 4800, Christchurch, New Zealand}
\author{J. A. Aguilar}
\affiliation{Universit{\'e} Libre de Bruxelles, Science Faculty CP230, B-1050 Brussels, Belgium}
\author{M. Ahlers}
\affiliation{Niels Bohr Institute, University of Copenhagen, DK-2100 Copenhagen, Denmark}
\author{M. Ahrens}
\affiliation{Oskar Klein Centre and Dept. of Physics, Stockholm University, SE-10691 Stockholm, Sweden}
\author{C. Alispach}
\affiliation{D{\'e}partement de physique nucl{\'e}aire et corpusculaire, Universit{\'e} de Gen{\`e}ve, CH-1211 Gen{\`e}ve, Switzerland}
\author{A. A. Alves Jr.}
\affiliation{Karlsruhe Institute of Technology, Institute for Astroparticle Physics, D-76021 Karlsruhe, Germany }
\author{N. M. Amin}
\affiliation{Bartol Research Institute and Dept. of Physics and Astronomy, University of Delaware, Newark, DE 19716, USA}
\author{K. Andeen}
\affiliation{Department of Physics, Marquette University, Milwaukee, WI, 53201, USA}
% ...
\maketitle

\email{analysis@icecube.wisc.edu}

\section{Introduction}
\label{sec:introduction}

At energies above \SI{40}{\tera \eV}, the Earth becomes opaque to neutrinos. For a power-law spectrum $\propto E^{-\gamma}$ at Earth's surface, the ratio of the flux arriving at IceCube to that at Earth's surface, $\fratio$, depends on the Earth column density, neutrino energy, $E_\nu$, spectral index $\gamma$, and neutrino cross section. The Earth column density is defined as $t(\theta)=\int_0^{y_{\rm {max}}} \rho(y, \theta) dy$, where $\theta$ is the arrival direction of the neutrino, $y_{\rm {max}}$ is its path length through the Earth, and $\rho(y, \theta)$ is the density at a point $y$ along the path. Figure~\ref{fig:attenuation2d} shows the electron neutrino and antineutrino $\fratio$ assuming a surface flux with $\gamma=2$. The spectral index affects the arrival flux through secondaries produced by tau decay in charged-current (CC) interactions or neutral-current (NC) interactions. In $\nu_e$ and $\nu_\mu$ CC interactions, the neutrino is effectively destroyed, whereas in $\nu_\tau$ CC interactions the outgoing tau-lepton may decay into lower-energy neutrinos~\cite{Halzen:1998be}. In NC interactions, the incoming neutrino is not destroyed but cascades down in energy~\cite{Formaggio:2013kya}. These flavor-dependent processes alter the neutrino flux as a function of the traversed path length~\cite{Vincent:2017svp}, which allows for probing the neutrino cross section at high energies. Finally, the dip in the $\bar{\nu}_e$ flux ratio due to the Glashow resonance~\cite{Glashow:1960zz} is visible in the right panel of Fig.~\ref{fig:attenuation2d} near $E_\nu = \SI{6.3}{\peta \eV}$. The Glashow resonance occurs from the interaction of an electron antineutrino with a bound atomic electron and is independent of the CC and NC interactions of nucleons.
\begin{figure*}[htp]
\centering
    \subfloat{
        \includegraphics[width=1\columnwidth]{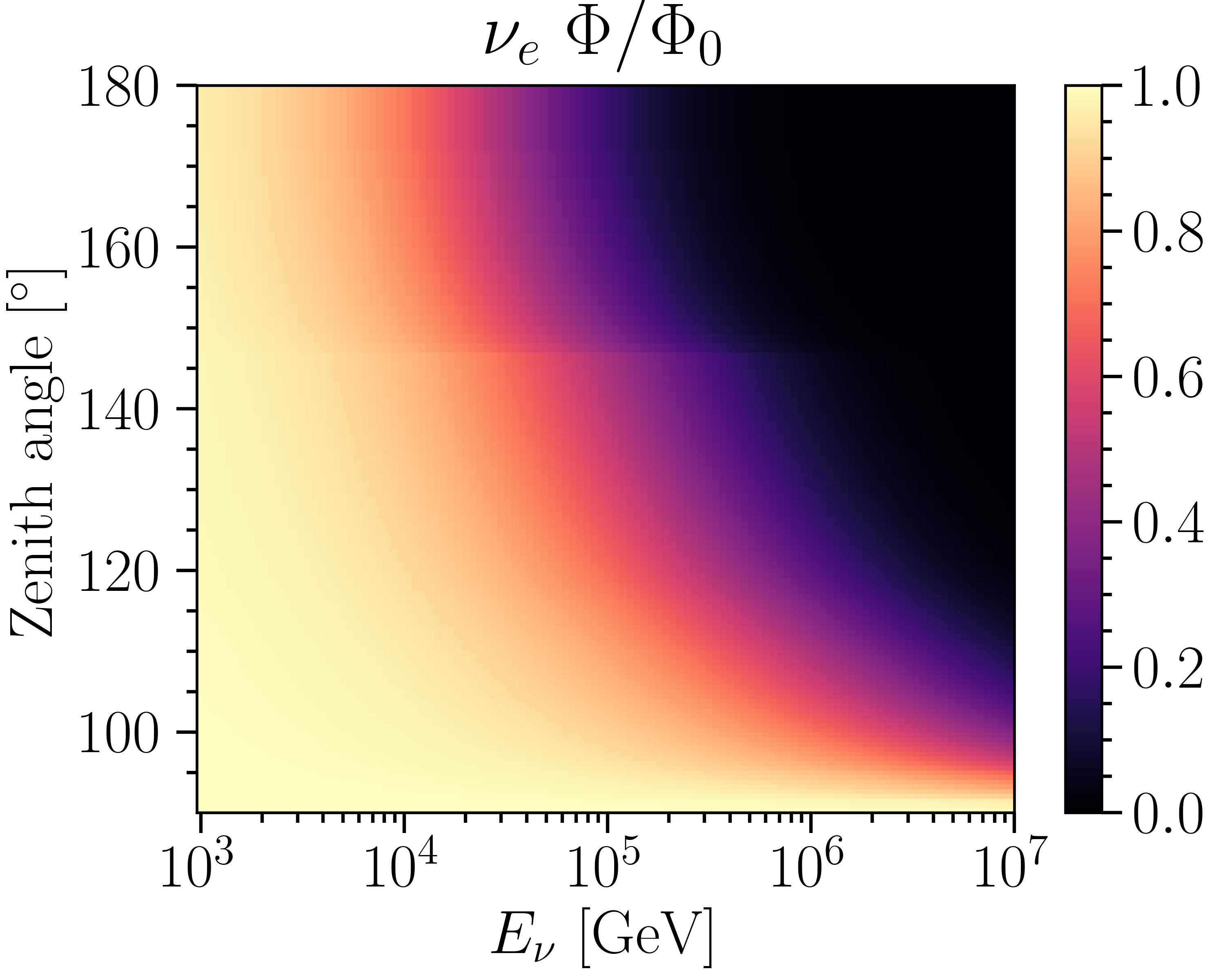}
    }
    \subfloat{
        \includegraphics[width=1\columnwidth]{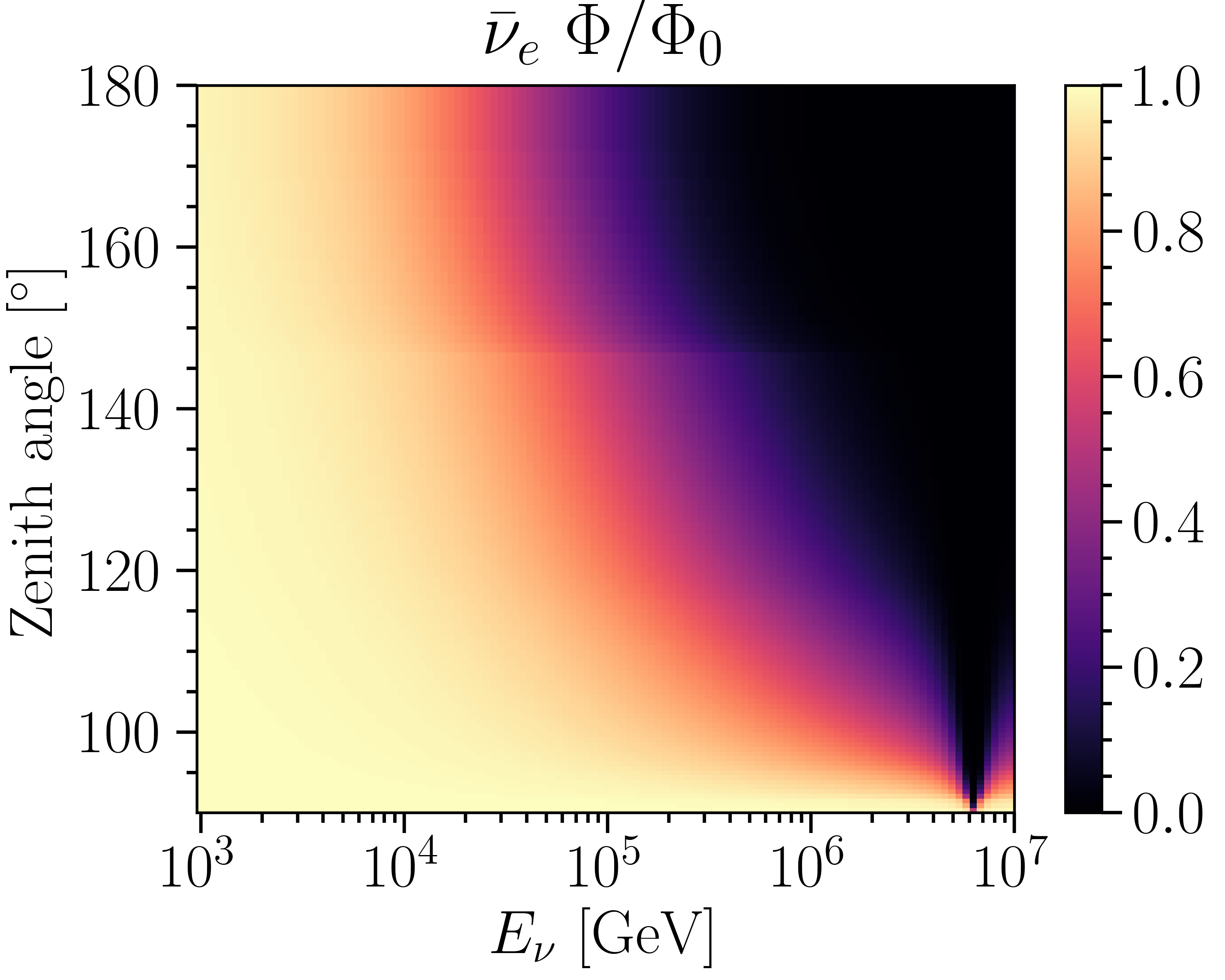}
    }
	\protect\caption{Ratio of the arrival flux to surface flux for both electron neutrinos and antineutrinos as a function of $E_\nu$ and zenith angle in IceCube detector coordinates. The flux at the surface is assumed to have a spectral index of $\gamma=2$. The core-mantle boundary is visible as a discontinuity near a zenith angle of \SI{147}{\degree}, and the enhanced suppression due to Glashow resonance is visible near \SI{6.3}{\peta \eV} in the electron antineutrino channel. Flux ratios for the other flavors are similar to that of electron neutrinos.}
    \label{fig:attenuation2d}
\end{figure*}
\iffalse
\begin{figure*}[htp]
\centering
    \subfloat{
        \includegraphics[width=0.3\textwidth]{fig/attenuation_nufate_nue}
    }
    \subfloat{
        \includegraphics[width=0.3\textwidth]{fig/attenuation_nufate_numu}
    }
    \subfloat{
        \includegraphics[width=0.3\textwidth]{fig/attenuation_nufate_nutau}
    }\\
    \subfloat{
        \includegraphics[width=0.3\textwidth]{fig/attenuation_nufate_nuebar}
    }
    \subfloat{
        \includegraphics[width=0.3\textwidth]{fig/attenuation_nufate_numubar}
    }
    \subfloat{
        \includegraphics[width=0.3\textwidth]{fig/attenuation_nufate_nutaubar}
    }
	\protect\caption{Ratio of the arrival flux to surface flux for both neutrinos and antineutrinos as a function of $E_\nu$ and zenith angle in IceCube detector coordinates. The flux at the surface is assumed to have a spectral index of $\gamma=2$.}
    \label{fig:attenuation2d}
\end{figure*}
\fi

The IceCube Neutrino Observatory, an in-ice neutrino detector situated at the South Pole, is capable of detecting high-energy neutrinos originating from both northern and southern hemispheres~\cite{Aartsen:2013bka, Aartsen:2013jdh, Aartsen:2014muf, Aartsen:2014gkd}. IceCube comprises over 5000 Digital Optical Modules (DOM) encompassing approxiately a cubic-kilometer of ice~\cite{Abbasi:2008aa, Abbasi:2010vc, Aartsen:2016nxy}. The ice acts as a detection medium by which Cherenkov radiation from charged particles produced in neutrino interactions can be observed. The high-energy starting events (HESE) sample selects events that interact within a fiducial region of the detector across a $4\pi$ solid angle~\cite{Aartsen:2014gkd, HESEPaper}. Here, we report a new cross-section measurement using information from all three neutrino flavors with 7.5 years of data.

In the Standard Model (SM), neutrino interactions are mediated by $W^\pm$ and $Z^0$ bosons for CC and NC channels, respectively. At energies above a few GeV, the dominant process is deep inelastic scattering (DIS) off of individual partons within the nucleon. Calculations in the perturbative QCD formalism rely on parton distribution functions (PDFs) obtained mostly from DIS experiments~\cite{Gandhi:1995tf, CooperSarkar:2011pa, Connolly:2011vc}. Uncertainties on the PDFs lead to uncertainties on the cross section. An alternative approach~\cite{Arguelles:2015wba} based on an empirical color dipole model of the nucleon along with the assumption that all cross sections increase at high energies as $\ln^2 s$ results in good agreement with the latest pQCD calculations. Proposed extensions of the SM based on large extra dimensions opening up above the Fermi scale predict a sharp rise in the neutrino-nucleon cross section above the SM value. One such model~\cite{Jain:2000pu} which was motivated by the claimed detection of cosmic rays above the GZK bound, assumes that neutrino-nucleon interaction is mediated by a massive spin-2 boson. This allows the neutrino-nucleon cross section to climb above \SI{e-27}{\cm^2} at $E_\nu > \SI{e19}{eV}$. Another possibility if spacetime has greater than four dimensions allows for the production of microscopic black holes in high-energy particle interactions and also leads to an increased neutrino-nucleon cross section above $\sim \SI{1}{\peta \eV}$~\cite{AlvarezMuniz:2002ga}. Such scenarios where the cross section increases steeply with energy could also be due to the existence of exotic particles such as leptoquarks~\cite{Romero:2009vu} or sphalerons~\cite{Ellis:2016dgb}, both of which have been discussed in the context of neutrino telescopes and could be probed via measurements of the high-energy neutrino cross section.

While SM calculations are generally consistent in the TeV-PeV energy range, few experimental measurements exist and none have been performed with all three neutrino flavors~\cite{Aartsen:2017kpd, Bustamante:2017xuy}. Recently, an IceCube measurement of the neutrino DIS cross section using up-going, muon neutrinos gave a result consistent with the Standard Model~\cite{Aartsen:2017kpd}. The measurement in~\cite{Bustamante:2017xuy} used showers in publicly available HESE data with six years of data-taking. This result, using the latest HESE sample with 7.5 years of data, includes classifiers for all three neutrino flavors and accounts for neutrinos from NC interactions and tau regeneration. Out of a total of 60 events above 60~TeV, 33 are also used in~\cite{Bustamante:2017xuy}. However, several updates described in~\cite{HESEPaper}, including the ice model~\cite{Chirkin:2014, Chirkin:2019vyq}, atmospheric neutrino passing fractions~\cite{Arguelles:2018awr}, likelihood construction~\cite{Arguelles:2019izp}, and systematics treatment~\cite{Aartsen:2017nmd}, affect their interpretation.

As the sample updates are detailed in~\cite{HESEPaper}, this paper focuses on the results of the neutrino-nucleon cross section measurement. A brief description of the event selection is given in Sec.~\ref{sec:sample}. Section \ref{sec:analysis} details the analysis procedure. Section \ref{sec:results} presents our Bayesian and frequentist results and compares them to existing measurements. We conclude in Sec.~\ref{sec:conclusions}.

\section{Event selection}
\label{sec:sample}

The measurements presented here rely on a sample of high-energy events that start within a fiducial region of the IceCube detector~\cite{Aartsen:2014gkd, HESEPaper}. In this context, events are taken to be the interaction by-products of neutrino interactions, or background muons from cosmic-ray interactions in the atmosphere. The \SI{90}{\meter} of the top and outer side layers of the detector, \SI{10}{\meter} of the bottom of the detector, and a \SI{60}{\meter} horizontal region near the highest concentration of dust in the ice are used as an active veto. Only events with fewer than 3 photoelectrons (PE) and fewer than 3 hit DOMs in the veto region within a predefined time window are kept. In addition, the total charge must exceed \SI{6000}{PE}~\cite{HESEPaper}. This removes almost all of the background due to atmospheric muons from the southern sky. Neutrinos arriving from above and below the detector are included in the sample, thus allowing for constraints across the full allowed region in zenith.  

Events are classified into three observable morphologies: cascades, tracks and double cascades. These classifiers are related to the true interaction channel of the neutrino. Electromagnetic and hadronic showers appear cascade-like, stochastic energy losses from high-energy muons appear track-like and the production and subsequent decay of a tau can appear as double cascades (in addition to the other two morphologies)~\cite{Usner:2018cel, HESEPaper}. Since a NC interaction produces a hadronic shower, it is not directly distinguishable from a CC interaction. In addition, misclassifications can occur and as such, the mapping from true to reconstructed observables is imperfect. To model such effects, detailed Monte-Carlo (MC) simulations are performed, taking into account systematic variations in the ice model. The MC is then processed in an identical manner as the data. It thus provides the connection from the physics parameters of interest to the observed data events.

\section{Analysis method}
\label{sec:analysis}

\begin{figure*}[htp]
\centering
    \subfloat{
        \includegraphics[width=1\columnwidth]{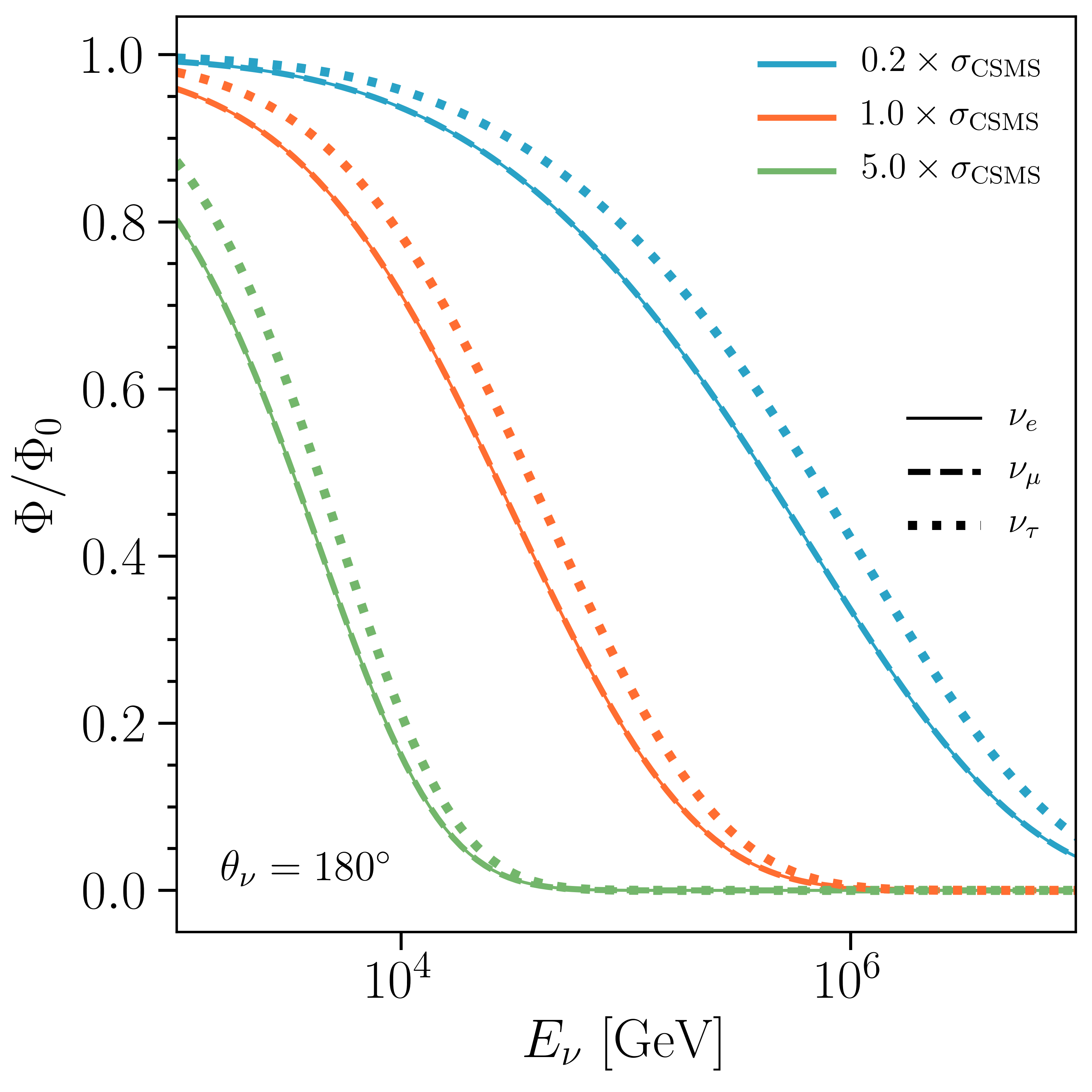}
    }
    \subfloat{
        \includegraphics[width=1\columnwidth]{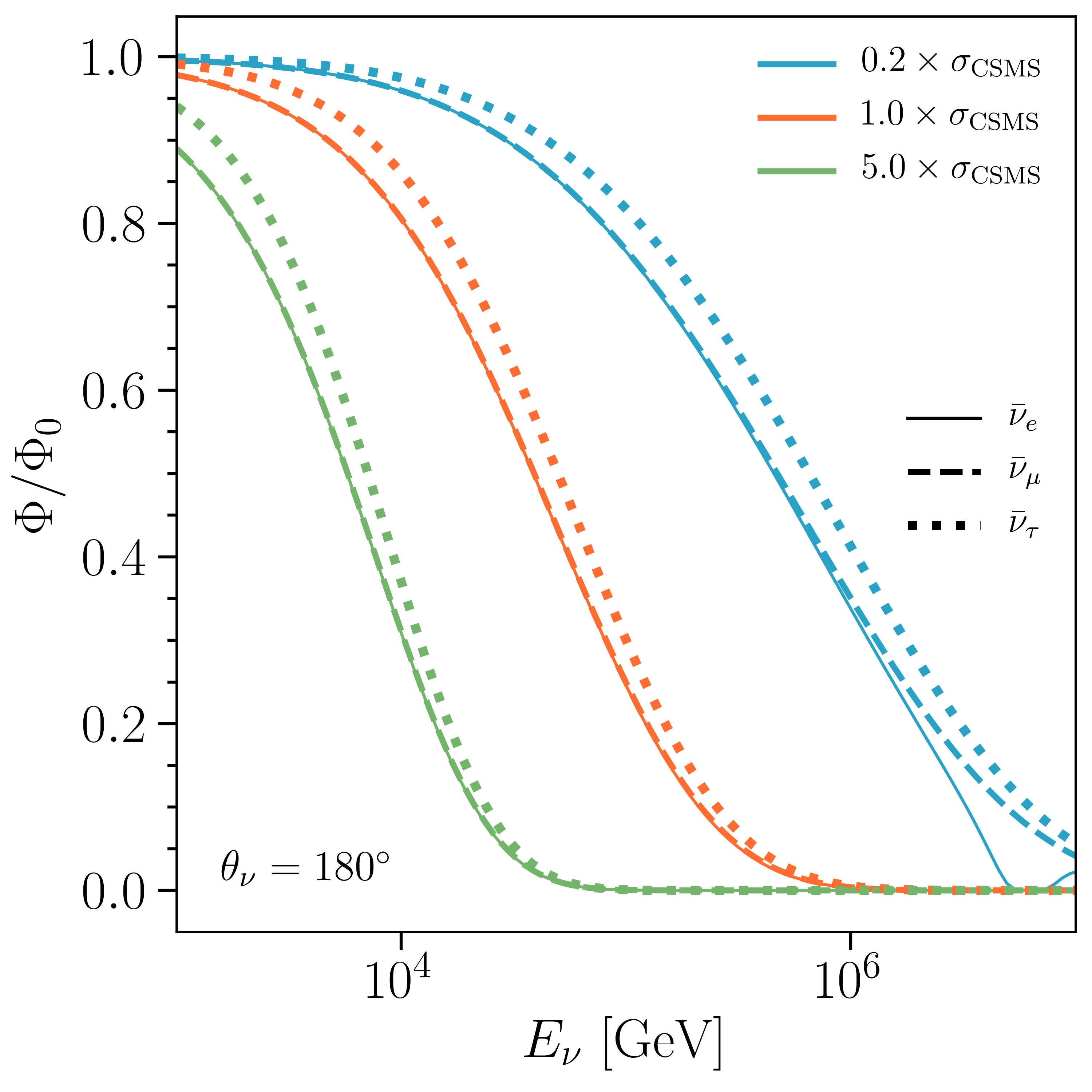}
    }

	\protect\caption{Ratio of the arrival flux to surface flux for both neutrinos and antineutrinos as a function of $E_\nu$ for three realizations of the cross section. The flux at the surface is assumed to have a spectral index of $\gamma=2$. The scaling is applied to the cross section given in~\cite{CooperSarkar:2011pa}.}
    \label{fig:attenuation1d}
\end{figure*}

Figure~\ref{fig:attenuation1d} illustrates the effect of scaling the DIS cross section up or down on the survival probability as a function of energy for a neutrino traveling through the full diameter of the Earth. It is plotted for each flavor individually as a function of the neutrino energy, $E_\nu$, at a zenith angle of $180^\circ$, and for a surface flux with spectral index of $\gamma=2$. The dependence on the spectrum arises from secondary neutrinos, which cascade down in energy, and are produced in NC interactions and tau decay~\cite{Halzen:1998be}. As the cross section increases, $\fratio$ decreases since the neutrinos are more likely to interact on their way through the Earth. The reason there is a slight flavor-dependence is due to the fact that CC $\barparen{\nu}_e$ and $\barparen{\nu}_\mu$ interactions are destructive, while a CC $\barparen{\nu}_\tau$ interaction produces a tau lepton which, unlike muons that lose most of their energy in the Earth before decaying due to their much longer lifetimes, can quickly decay to a lower-energy $\nu_\tau$. Neutral current interactions have a similar effect, and these effects are taken into account~\cite{Aartsen:2017kpd, Vincent:2017svp}. Furthermore, the dip in the $\bar{\nu}_e$ flux ratio due to the Glashow resonance is again visible. This effect, taken over the full 2D energy-zenith distribution, allows us to place constraints on the cross section itself.

In this paper, we report the neutrino DIS cross section as a function of energy under a single-power-law astrophysical flux assumption.  Four scaling parameters, $\bm{x}=(x_0, x_1, x_2, x_3)$, are applied to the cross section given in~\cite{CooperSarkar:2011pa} (CSMS) across four energy bins with edges fixed at 60~TeV, 100~TeV, 200~TeV, 500~TeV, and 10~PeV, where the indexes correspond to the ordering of the energy bins from lowest to highest energies. Each parameter linearly scales the neutrino and antineutrino DIS cross section in each bin, while keeping the ratio of CC-to-NC contributions fixed. The fixed CC-to-NC ratio implies that this analysis should not be interpreted as a direct test of the large extra dimensions model~\cite{Jain:2000pu}, which only applies to NC interactions. At energies above 1~TeV, the neutrino-nucleon cross sections for all three neutrino flavors converge. The cross section is therefore assumed to not depend on flavor in this measurement, but any differences in the arrival flux of $\nu_e$, $\nu_\mu$, $\nu_\tau$ are taken into account. As the cross section is not flat in each bin, the effect of these four parameters is to convert it into a piece-wise function where each piece is independently rescaled. Such an approach introduces discontinuities due to binning, but allows for a measurement of the total neutrino-nucleon cross section as a function of energy. It also relaxes constraints based on the overall shape of the CSMS cross section and results in a more model-independent measurement. As the fit proceeds over all four bins simultaneously, bin-to-bin correlations can be examined, though no regularization is applied.

The CSMS cross section is computed for free nucleon targets and does not correct for nuclear shadowing. The shadowing effect modifies nuclear parton densities and is stronger for heavier nuclei. At energies below \SI{100}{\tera \eV} antishadowing can increase the cross section by 1--2\,\% , while above \SI{100}{\tera \eV} shadowing can decrease the cross section by 3--4\,\%~\cite{Klein:2020nuk}. As this is a subdominant effect, we do not include it in this analysis. We do, however, consider the Glashow resonance in which an incident $\bar{\nu_e}$ creates an on-shell $W^-$ by scattering off an electron in the detector. 

The effect on the expected arrival flux at the detector due to a modified cross section is calculated with \nusquids{}, which properly takes into account destructive CC interactions, cascading NC interactions, and tau-regeneration effects~\cite{Delgado:2014kpa}. For each $\bm{x}$, neutrino events in MC are reweighted by $x_i \Phi(E_\nu, \theta_\nu, \bm{x})/\Phi(E_\nu, \theta_\nu, \bm{1})$, where $\Phi$ is the arrival flux as calculated by \nusquids{}, $E_\nu$ is the true neutrino energy, $\theta_\nu$ the true neutrino zenith angle, and $x_i$ the cross section scaling factor at $E_\nu$. A forward-folded fit is then performed, relying on MC to map each neutrino flavor to the experimental particle identification (PID) of tracks, cascades and double cascades~\cite{Usner:2018cel}, in the reconstructed zenith vs reconstructed energy distribution for tracks and cascades, and in the reconstructed energy vs cascade length separation distribution for double cascades~\cite{HESEPaper}. The fit uses the Poisson-like likelihood, $\leff$, which accounts for statistical uncertainties in the MC and is constructed by comparing the binned MC to data~\cite{Arguelles:2019izp}. The ternary PID of tracks, cascades, and double cascades is an additional constraint to the fit which allows this measurement to incorporate interaction characteristics of all three neutrino flavors~\cite{Usner:2018cel}. Using MC simulations, we can account for deviations between the true flavor and the PID, and also estimate its accuracy. Under best-fit expectations, true $\nu_e$ are classified as cascades $\sim 57 \%$ of the time, true $\nu_\mu$ as tracks $\sim 73 \%$ of the time, and true $\nu_\tau$ as double cascades $\sim 65 \%$ of the time~\cite{HESEPaper}.

Systematic uncertainties on the atmospheric neutrino flux normalization where the neutrinos are produced by $\pi$ or $K$ decay~\cite{Honda:2006qj}, $\Phi_\texttt{conv}$, atmospheric neutrino flux normalization where the neutrinos are produced by charm meson decay~\cite{Bhattacharya:2015jpa}, $\Phi_\texttt{prompt}$, astrophysical spectral index, $\gamma$, astrophysical flux normalization, $\Phi_\texttt{astro}$, atmospheric muon flux normalization, $\Phi_{\mu}$, $\pi/K$ ratio, atmospheric $\nu/\bar{\nu}$ ratio, and the cosmic ray spectral index~\cite{Gaisser:2013bla}, $\delta \gamma_\texttt{CR}$, are taken into account. Detector systematic studies were performed using Asimov data but had a negligible impact on the result. Priors on the nuisance parameters are given in Table~\ref{tab:nuisances}. The prior on $\gamma_\texttt{astro}$ is driven by the usual Fermi acceleration mechanism, allowing for a large uncertainty that covers those reported in a previous and independent IceCube measurement of the diffuse neutrino flux~\cite{Aartsen:2016xlq}. Such a large uncertainty minimizes the impact of changing the central value on the measured cross section. None of the $x_i$ parameters shifted by more than \SI{1}{\percent} in post-unblinding checks where $\gamma_\texttt{astro}=3.0 \pm 1.0$.

Out of all the nuisance parameters, $\gamma_\texttt{astro}$ and $\Phi\texttt{astro}$ exhibited the largest correlation with the cross section parameters. They are most strongly correlated with $x_0$, the cross section in the lowest-energy bin. This is believed to be related to the fact that lower-energy neutrinos are subject to less Earth-absorption so the main effect of varying the low-energy cross section is a near-linear scaling at the detector. This makes $x_0$ essentially inversely proportional to the astrophysical flux. By allowing the cross section to float the data seems to prefer the softer index, as given in Table~\ref{tab:nuisances}. 

\begin{table}[thb]
\centering
\begin{tabular}{l|rrrr}
Parameter & Constr./Prior  & Range & Shape & Best fit\\ 
\hline
\multicolumn{1}{l|}{\textbf{Astro.~$\nu$:}} & & & &\\
$\Phi_\texttt{astro}$ & - & $[0,\infty)$ & Uniform & 6.94\\
$\gamma_\texttt{astro}$ & $2.0\pm1.0$ &  $(-\infty,\infty)$ & Gaussian & 3.15\\
&&&&\\
\multicolumn{1}{l|}{\textbf{Atmos.~$\nu$:}} & & & &\\
$\Phi_\texttt{conv}$ & $1.0\pm0.4$ & $[0, \infty)$ & Truncated & 0.96\\
$\Phi_\texttt{prompt}$ & $1.0\pm3.0$ & $[0, \infty)$ & Truncated & 0.00\\
$\pi/K$ & $1.0\pm0.1$ & $[0, \infty)$ & Truncated & 1.00\\
${2\nu/\left(\nu+\bar{\nu}\right)}_\texttt{atmo}$ & $1.0\pm0.1$ & $[0,2]$ & Truncated & 1.00 \\
&&&&\\
\multicolumn{1}{l|}{\textbf{Cosmic-ray:}} & & & &\\
$\Delta\gamma_\texttt{CR}$ & $-0.05\pm 0.05$ & $(-\infty,\infty)$ & Gaussian& -0.05\\
$\Phi_\mu$ & $1.0\pm 0.5$ & $[0,\infty)$ & Truncated & 1.22\\
\end{tabular}
\caption{Central values and uncertainties on the nuisance parameters included in the fit. Truncated Gaussians are set to zero outside the range. These modify the likelihood used in both the Bayesian and frequentist constructions. Their best-fit values over the likelihood space are also given.}
\label{tab:nuisances}
\end{table}

The interaction rate of high-energy neutrinos traveling through the Earth is also dependent on the Earth density. Here, we fix the density to the preliminary reference Earth model (PREM)~\cite{Dziewonski:1981xy}. This is a parametric description of the density as a function of radial distance from the center of the Earth, evaluated using several sources of surface and body seismic wave data. Since the density uncertainty is at the few percent level, it is negligible in comparison to the flux uncertainty and is fixed for the purposes of this measurement~\cite{Kennett:1998}.

Note that the Glashow resonance occurs for an incident $\bar{\nu_e}$ with an energy around \SI{6.3}{PeV} and is not varied in the fit as it is calculable from first principles, using the known decay width of the $W$ boson. However, unlike high-energy neutrino-nucleon scattering, the expected number of events due to the Glashow resonance is strongly dependent on the ratio of neutrinos and antineutrinos in the incident flux. We therefore performed a test that varied the astrophysical flux from a pure neutrino flux to a pure antineutrino flux. It was found only to have a minimal effect in the highest energy bin, where the measurement uncertainty is largest. This is due in part to the steeply falling spectrum, which causes the flux at \SI{6.3}{PeV} to be much smaller than that at lower energies. As the effect on the cross section is minimal, we keep the ratio of the flux of astrophysical neutrinos and antineutrinos fixed to unity.

We report both Bayesian highest posterior density (HPD) credible intervals and frequentist confidence intervals (CI). In the Bayesian construction, the posterior on the four scaling parameters are obtained with a MCMC sampler, \emcee, marginalizing over nuisance parameters~\cite{ForemanMackey:2012ig}. A uniform prior from 0 to 50 is assumed for all four cross section scaling parameters. Such a prior gives more weight to parameter values greater than one. To test its effect, the MCMC was also run assuming a log-uniform prior which gives a results consistent with those assuming a uniform prior. The MCMC is sampled with 60 walkers over 5000 total steps, the first 1000 of which are treated as part of the initialization stage and discarded.

The frequentist confidence regions are obtained from a grid scan of the likelihood across four dimensions, profiling over the nuisance parameters and assuming Wilks' theorem.  For $x_0$, $x_1$, and $x_2$, 15 equal-distant points are used from 0.1--5. For $x_3$, 29 equal-distant points are used from 0.1--9.9. For each $\bm{x}$ on the mesh of these points, the likelihood is minimized over all other nuisance parameters. Confidence regions in two or one dimension are then evaluated by profiling across the other cross-section parameters followed by application of Wilks' theorem. Though the best-fit $\Phi_\texttt{prompt}=0$, the prompt component is expected to be a small contribution to the overall distribution. Thus we expect Wilks' theorem to hold asymptotically in the high statistics limit.

\begin{figure}[htp]
    \includegraphics[width=1\columnwidth]{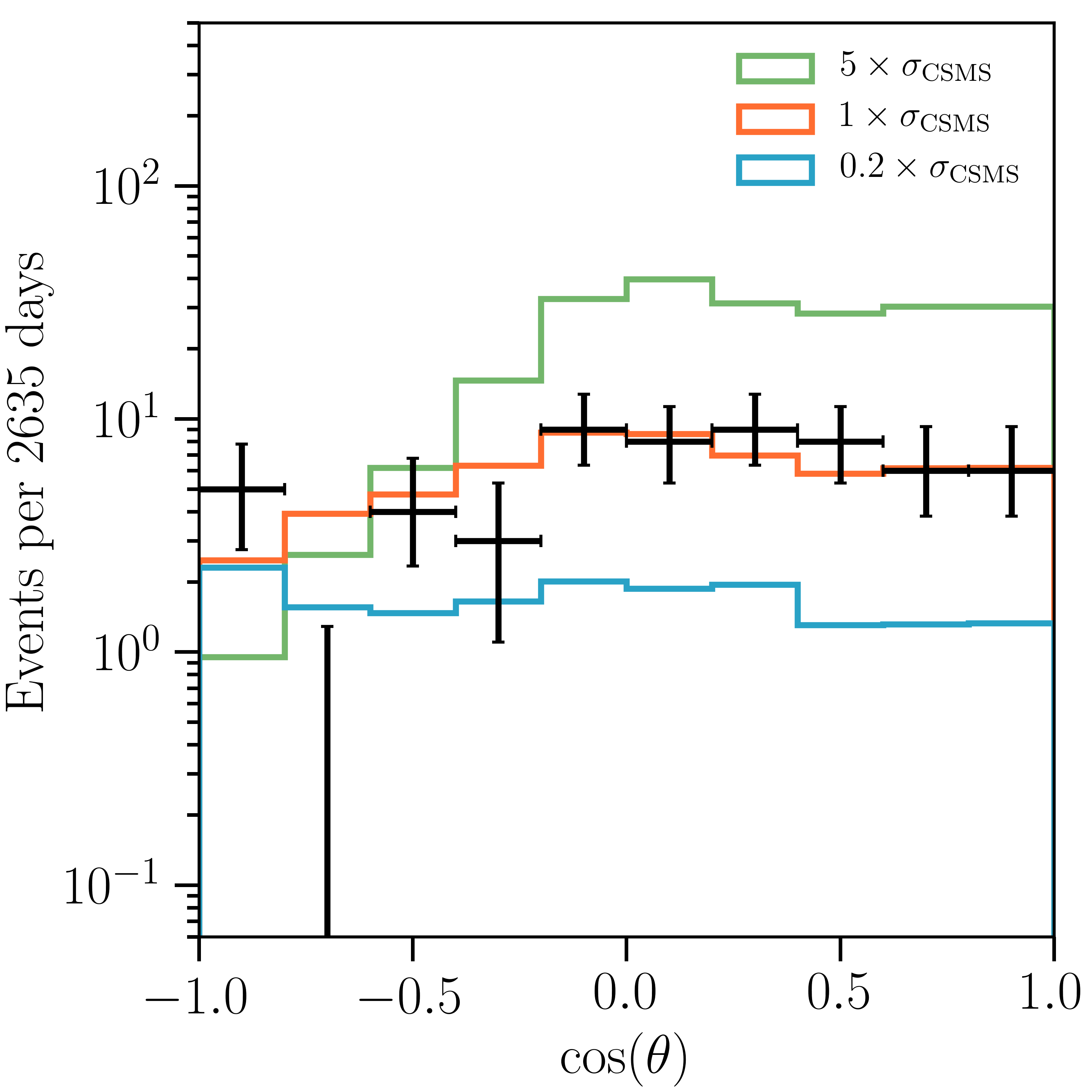}
	\protect\caption{The zenith distribution of data and the best-fit, single-power-law flux expectation assuming $\sigma_{\rm CSMS}$ (orange)~\cite{CooperSarkar:2011pa}. Predictions from two alternative cross sections are shown as well, assuming the same flux. In the southern sky, $\cos \theta > 0$, the Earth absorption is negligible so the effect of rescaling the cross section is linear. In the northern sky, $\cos \theta < 0$, the strength of Earth absorption is dependent on the cross section, as well as the neutrino energy and zenith angle.}
    \label{fig:xzen}
\end{figure}

The zenith-dependent effect of the cross section on the event rate is shown in Fig.~\ref{fig:xzen}, assuming the best-fit, single-power-law flux reported in~\cite{HESEPaper}, which is obtained using the CSMS cross section $\sigma=\sigma_{\rm CSMS}$~\cite{CooperSarkar:2011pa}. The degeneracy in the measurements of flux and cross section is broken by the different amounts of matter traversed by neutrinos arriving from different directions. In order to illustrate the effect of a modified cross section, two alternative expectations are shown for $\sigma=0.2 \sigma_{\rm CSMS}$ and $\sigma=5 \sigma_{\rm CSMS}$ under the same best-fit flux assumption. In the southern sky ($\cos \theta > 0$) the Earth absorption is negligible and the event rate is simply proportional to the cross section. In the northern sky ($\cos \theta < 0$) the strength of Earth absorption is dependent on the zenith angle and $E_\nu$, as shown in Fig.~\ref{fig:attenuation2d}, as well as the cross section, shown for a single zenith angle in Fig.~\ref{fig:attenuation1d}. Absorption alters the shape of the event-rate zenith distribution in the northern sky. For example, with $\sigma= 5 \sigma_\mathrm{CSMS}$ and near $\cos \theta = -0.5$, the attenuation of the arriving flux counteracts the increased neutrino interaction probability, so that the event rate falls back to that expected from the CSMS cross section. Modifications of the neutrino cross section are thus constrained by the non-observation of energy-dependent distortions in the zenith angle distribution.

\section{Results}
\label{sec:results}

\begin{figure}[htpb]
    \includegraphics[width=1\columnwidth]{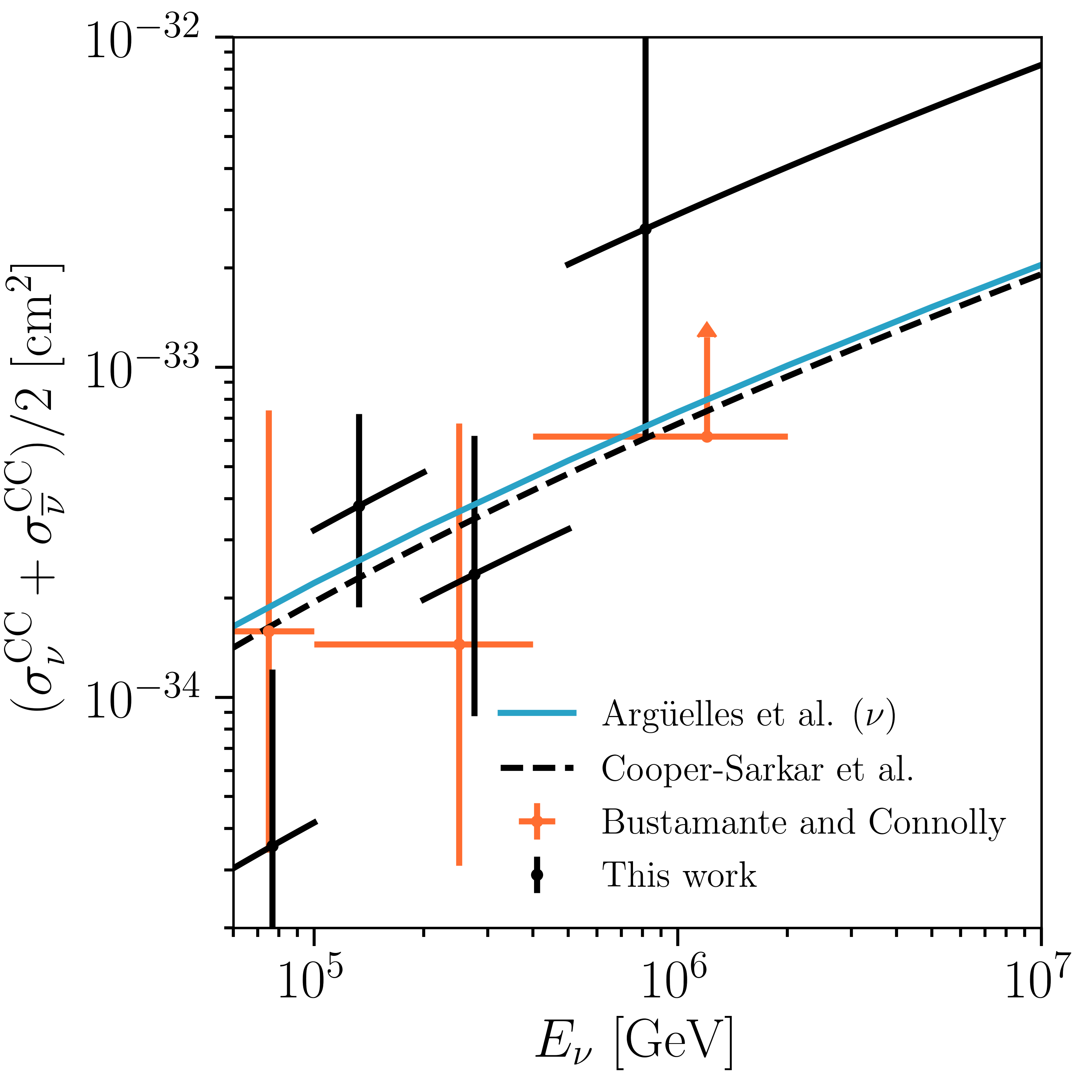}  
	\protect\caption{The charged-current, high-energy neutrino cross section as a function of energy, averaged over $\nu$ and $\bar{\nu}$. The Bayesian 68.3\% HPD credible interval is shown along with two cross section calculations~\cite{CooperSarkar:2011pa, Arguelles:2015wba}. The credible intervals from a previous analysis~\cite{Bustamante:2017xuy} are also shown for comparison.}
    \label{fig:xsmcmc}
\end{figure}

\begin{figure}[htpb]
    \includegraphics[width=1\columnwidth]{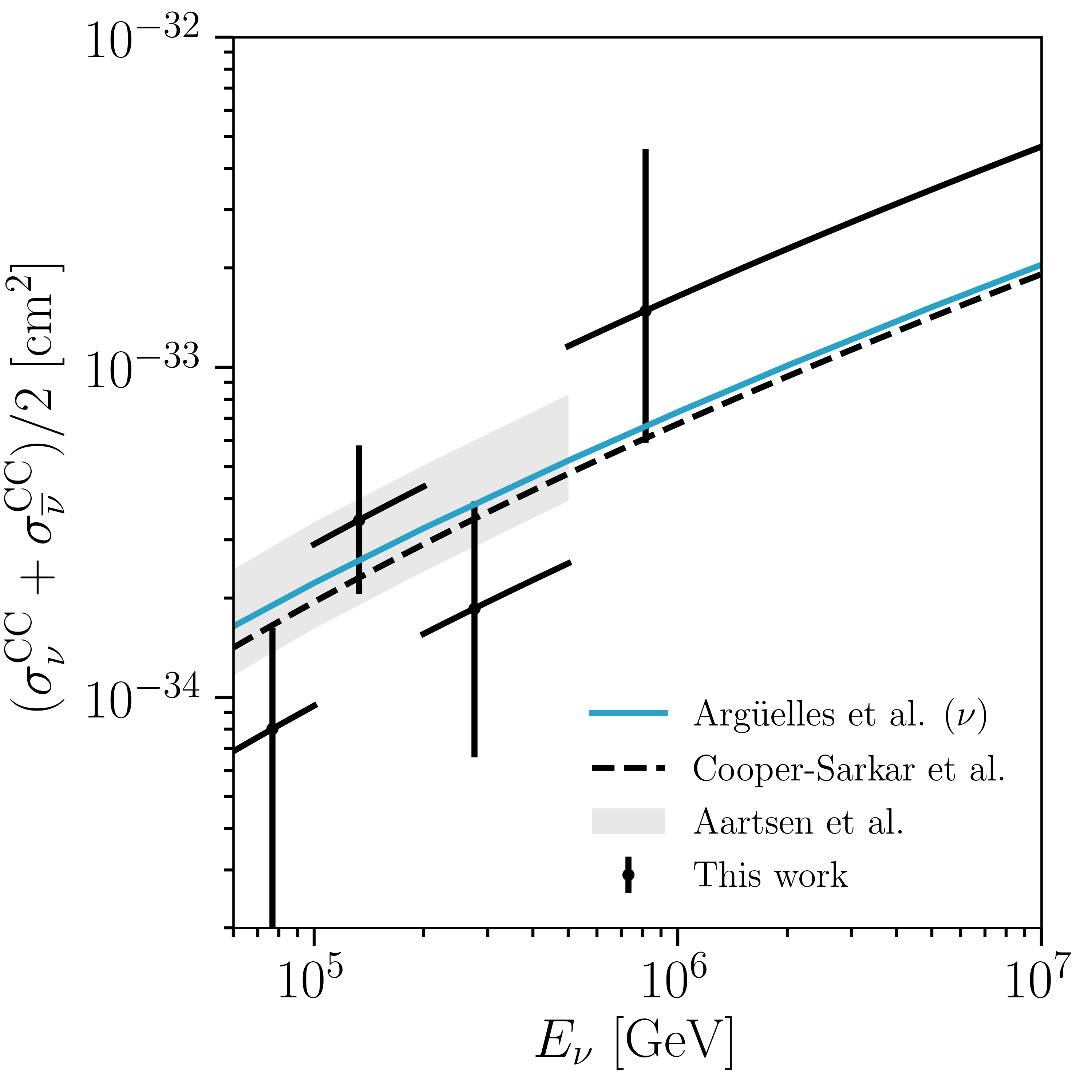}  
	\protect\caption{The charged-current, high-energy neutrino cross section as a function of energy, averaged over $\nu$ and $\bar{\nu}$. The Wilks' 1-sigma CI is shown along with two cross section calculations~\cite{CooperSarkar:2011pa, Arguelles:2015wba}. The confidence intervals from~\cite{Aartsen:2017kpd} are also shown for comparison}
    \label{fig:xsfreq}
\end{figure}

The CC cross section, averaged over $\nu$ and $\overline{\nu}$, are shown in black in Fig.~\ref{fig:xsmcmc} and Fig.~\ref{fig:xsfreq} for the Bayesian 68.3\% HPD and frequentist one sigma intervals assuming Wilks' theorem, respectively. As the scale factor is applied across the entire interval within an energy bin on the CSMS calculations, the shape is preserved within each bin. The central point in each energy bin corresponds to the expected, most-probable energy in $dN_{MC}/d\log E$, the distribution of events in the MC along the x-axis. This is chosen in lieu of the linear or logarithmic bin center to better represent where most of the statistical power lies in each bin. Since we assume a fixed CC-NC cross-section ratio, the NC cross section is the same result relative to the CSMS prediction and so is not shown here.

In addition, the measurement based on HESE showers with six years of data is shown as orange crosses~\cite{Bustamante:2017xuy} in Fig.~\ref{fig:xsmcmc} and the previously published IceCube measurement, using upgoing muon-neutrinos, is shown as the shaded gray region~\cite{Aartsen:2017kpd} in Fig.~\ref{fig:xsfreq}. Since credible intervals and confidence intervals have different interpretations, we do not plot them on the same figure. Note that both previous measurements extend below 60~TeV and are truncated in this comparison. Predictions from~\cite{CooperSarkar:2011pa} and~\cite{Arguelles:2015wba} are shown as the dashed and solid lines, respectively.

\begin{figure}[htp]
    \includegraphics[width=1\columnwidth]{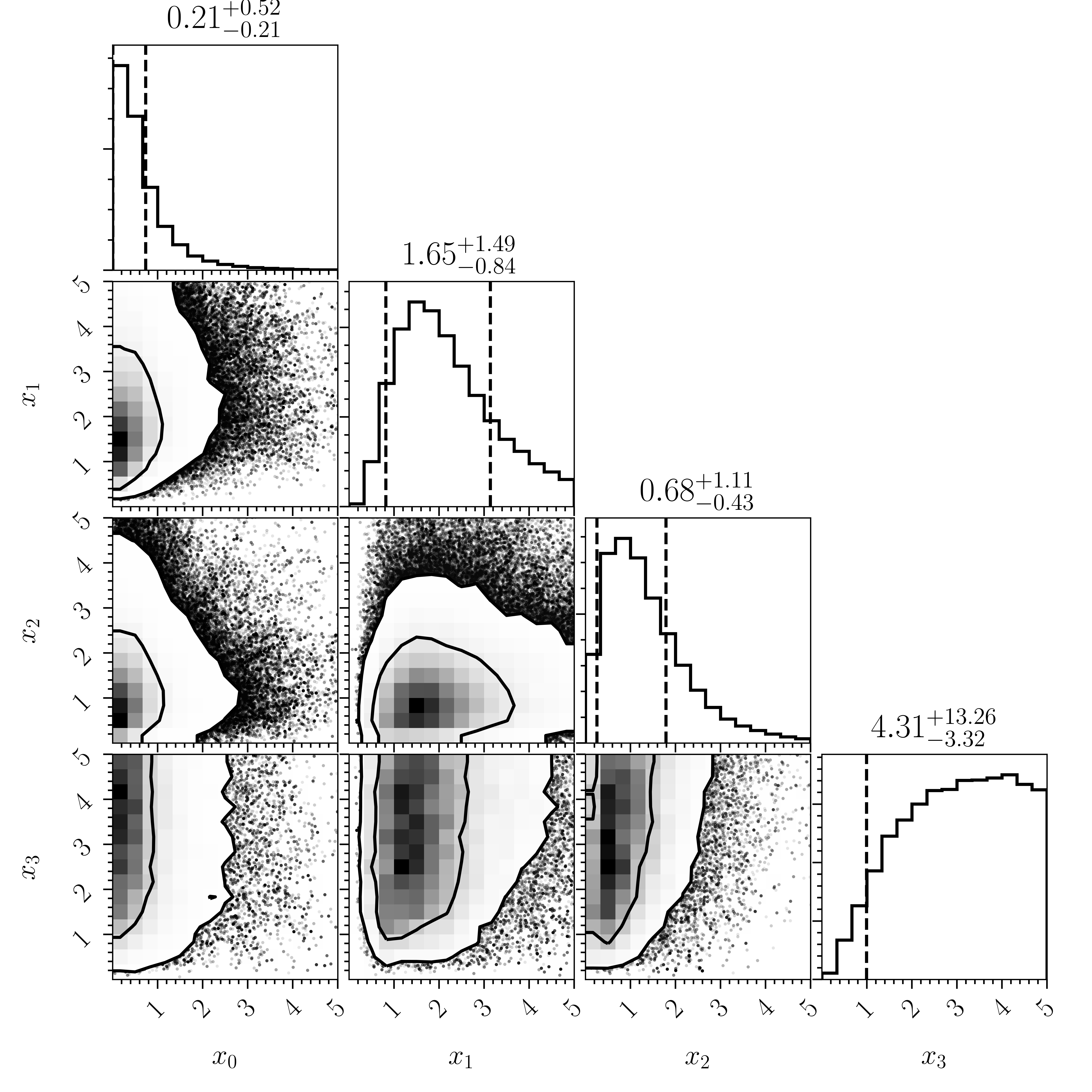}  
	\protect\caption{The full posterior distribution of $\bm{x}$ as evaluated with \emcee~\cite{ForemanMackey:2012ig}. In the two-dimensional distributions, the $68.3\%$ and $95.4\%$ HPD regions are shown. In the one-dimensional distribution, the $68.3\%$ HPD interval is indicated by the dashed lines.}
    \label{fig:corner}
\end{figure}

\begin{figure}[htp]
    \includegraphics[width=\columnwidth]{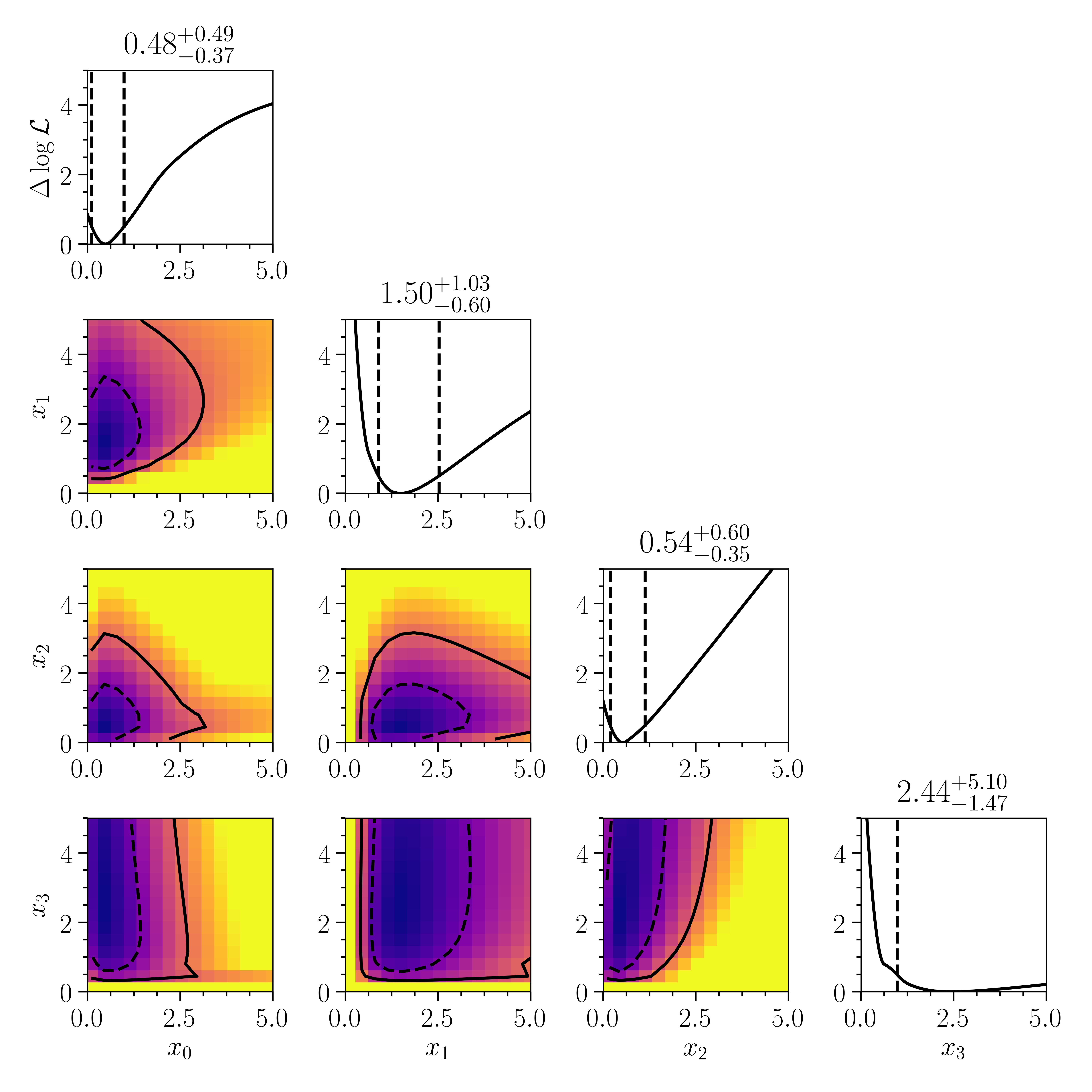}  
	\protect\caption{The profile likelihood of $\bm{x}$ as evaluated with the grid scan over $\bm{x}$. In the two-dimensional figures, the Wilks' $68.3\%$ and $95.4\%$ confidence regions are shown as dashed and solid lines, respectively. In the one-dimensional plots of $\Delta \log \mathcal{L}$, the $68.3\%$ confidence interval is indicated by the dashed lines.}
    \label{fig:freq}
\end{figure}

A corner plot of the posterior density, marginalized over all except two or one of the cross-section parameters, is shown in Fig.~\ref{fig:corner}. Similarly, two-dimensional profile likelihoods are shown in Fig.~\ref{fig:freq}. Both exhibit little correlation between the various cross-section parameters. The largest uncertainty arises for $x_3$, which has the widest posterior distribution and flattest profile likelihood.

The Bayesian and frequentist results are consistent with each other, though again we caution that their intervals cannot be interpreted in the same manner. The results are compatible with the Standard Model and are summarized in Table~\ref{tab:results}.
\begin{table}
    \centering
    \begin{tabular}{l|rrr}
    Parameter & Energy range & 68.3\% HPD & 68.3\% CI\\
    \hline
    $x_0$ & \SIrange{60}{100}{\tera \eV}& $0.21^{+0.52}_{-0.21}$ & $0.48^{+0.49}_{-0.37}$\\
    $x_1$ & \SIrange{100}{200}{\tera \eV}& $1.65^{+1.49}_{-0.84}$ & $1.50^{+1.03}_{-0.60}$\\
    $x_2$ & \SIrange{200}{500}{\tera \eV}& $0.68^{+1.11}_{-0.43}$ & $0.54^{+0.60}_{-0.35}$\\
    $x_3$ & \SIrange[exponent-to-prefix = true,scientific-notation = engineering,zero-decimal-to-integer]{500}{10000}{\tera \eV}& $4.31^{+13.26}_{-3.32}$ & $2.44^{+5.10}_{-1.47}$\\
    \end{tabular}
    \caption{Measured 68.3\% HPD (Bayesian) and CI (frequentist) for the four cross section parameters.}
    \label{tab:results}
\end{table}

\section{Conclusions}
\label{sec:conclusions}

We have described a measurement of the neutrino DIS cross section using the IceCube detector. Variations in the neutrino cross section from Standard Model predictions modify the expected flux and event rate at our detector, and a sample of high-energy events starting within the fiducial volume of IceCube has been utilized to thus measure the neutrino cross section. Previous TeV-PeV scale neutrino cross sections have been measured by IceCube~\cite{Aartsen:2017kpd} using a sample of throughgoing muons, and with cascades in the HESE sample~\cite{Bustamante:2017xuy}. This result, however, is the first measurement of the neutrino DIS cross section to combine information from all three neutrino flavors.

Our results are compatible with Standard Model predictions, though the data seems to prefer smaller values at the lowest-energy bin, and higher values at the highest-energy bin. There does not seem to be strong correlations between the cross section bins, though large uncertainties due to a dearth of data statistics make it difficult to draw strong conclusions. With additional data, or with a combined fit across multiple samples, more precise measurements are foreseen in the near future~\cite{Robertson:2019wfw}.

\begin{acknowledgements}
The IceCube collaboration acknowledges the significant contributions to this manuscript from Tianlu Yuan. The authors gratefully acknowledge the support from the following agencies and institutions: USA {\textendash} U.S. National Science Foundation-Office of Polar Programs,
U.S. National Science Foundation-Physics Division,
Wisconsin Alumni Research Foundation,
Center for High Throughput Computing (CHTC) at the University of Wisconsin{\textendash}Madison,
Open Science Grid (OSG),
Extreme Science and Engineering Discovery Environment (XSEDE),
U.S. Department of Energy-National Energy Research Scientific Computing Center,
Particle astrophysics research computing center at the University of Maryland,
Institute for Cyber-Enabled Research at Michigan State University,
and Astroparticle physics computational facility at Marquette University;
Belgium {\textendash} Funds for Scientific Research (FRS-FNRS and FWO),
FWO Odysseus and Big Science programmes,
and Belgian Federal Science Policy Office (Belspo);
Germany {\textendash} Bundesministerium f{\"u}r Bildung und Forschung (BMBF),
Deutsche Forschungsgemeinschaft (DFG),
Helmholtz Alliance for Astroparticle Physics (HAP),
Initiative and Networking Fund of the Helmholtz Association,
Deutsches Elektronen Synchrotron (DESY),
and High Performance Computing cluster of the RWTH Aachen;
Sweden {\textendash} Swedish Research Council,
Swedish Polar Research Secretariat,
Swedish National Infrastructure for Computing (SNIC),
and Knut and Alice Wallenberg Foundation;
Australia {\textendash} Australian Research Council;
Canada {\textendash} Natural Sciences and Engineering Research Council of Canada,
Calcul Qu{\'e}bec, Compute Ontario, Canada Foundation for Innovation, WestGrid, and Compute Canada;
Denmark {\textendash} Villum Fonden, Danish National Research Foundation (DNRF), Carlsberg Foundation;
New Zealand {\textendash} Marsden Fund;
Japan {\textendash} Japan Society for Promotion of Science (JSPS)
and Institute for Global Prominent Research (IGPR) of Chiba University;
Korea {\textendash} National Research Foundation of Korea (NRF);
Switzerland {\textendash} Swiss National Science Foundation (SNSF);
United Kingdom {\textendash} Department of Physics, University of Oxford.
\end{acknowledgements}

\bibliography{xs}

\end{document}